\pdfoutput=1
\documentclass[3p,times,twocolumn,sort&compress]{elsarticle}

\usepackage{ecrc-ac}
\usepackage[british]{babel}

\volume{00}

\firstpage{1}

\journalname{Nuclear Physics B Proceedings Supplement}

\runauth{M. Steinhauser, T. Ueda, J.A.M. Vermaseren}

\jid{nuphbp}

\jnltitlelogo{Nuclear Physics B Proceedings Supplement}

\usepackage{amssymb}
\usepackage[figuresright]{rotating}

\newcommand{\form}{{\tt FORM}}
\newcommand{\parform}{{\tt ParFORM}}
\newcommand{\tform}{{\tt TFORM}}

\begin{document}

\begin{frontmatter}

%- {{{ Title, authors and addresses:

%% use the tnoteref command within \title for footnotes;
%% use the tnotetext command for the associated footnote;
%% use the fnref command within \author or \address for footnotes;
%% use the fntext command for the associated footnote;
%% use the corref command within \author for corresponding author footnotes;
%% use the cortext command for the associated footnote;
%% use the ead command for the email address,
%% and the form \ead[url] for the home page:
%%
%% \title{Title\tnoteref{label1}}
%% \tnotetext[label1]{}
%% \author{Name\corref{cor1}\fnref{label2}}
%% \ead{email address}
%% \ead[url]{home page}
%% \fntext[label2]{}
%% \cortext[cor1]{}
%% \address{Address\fnref{label3}}
%% \fntext[label3]{}

\dochead{}
%% Use \dochead if there is an article header, e.g. \dochead{Short communication}

\title{Parallel versions of \form{} and more}

%% use optional labels to link authors explicitly to addresses:
%% \author[label1,label2]{<author name>}
%% \address[label1]{<address>}
%% \address[label2]{<address>}

\author{Matthias Steinhauser, 
  Takahiro Ueda}

\address{Institut f{\"u}r Theoretische Teilchenphysik, Karlsruhe
  Institute of Technology (KIT), 76128 Karlsruhe, Germany}

\author{Jos A.M. Vermaseren}

\address{Nikhef, Science Park 105, 1098 XG Amsterdam, The Netherlands}

\begin{abstract}
  We review the status of the parallel versions of the computer algebra system
  \form{}. In particular, we provide a brief overview about the historical
  developments, discuss the strengths of \parform{} and \tform{}, and
  mention typical applications.  Furthermore, we briefly discuss the programs
  {\tt FIRE} and {\tt FIESTA}, which have also been developed with the
  Collaborative Research Center/TR~9 (CRC/TR~9).
\end{abstract}

\begin{keyword}
  Computer algebra \sep \form{} \sep multi-loop integrals \sep reduction to
  master integrals \sep numerical evaluation of Feynman integrals
%% keywords here, in the form: keyword \sep keyword

%% MSC codes here, in the form: \MSC code \sep code
%% or \MSC[2008] code \sep code (2000 is the default)

\end{keyword}

%- }}}

\end{frontmatter}

\graphicspath{ {figs_a2/} }

%%
%% Start line numbering here if you want
%%
% \linenumbers

%% main text

%- {{{ Introduction:

\section{\label{sec::intro}Introduction}

The symbolic manipulation of complicated formulae has a long tradition in 
particle physics. Computer algebra systems (CAS) have been used already 
quite early in order to evaluate, e.g., traces over $\gamma$ matrices. 
Among the first CAS there are {\tt REDUCE}~\cite{Hearn:1971zza} by
A.~Hearn, {\tt SCHOONSCHIP}~\cite{Schoonschip,Schoonschip2,Veltman:1991xb},
designed by M.~Veltman, {\tt ASHMEDAI}~\cite{ashmedai} by M.~Levine,
and {\tt Macsyma}~\cite{macsyma} developed at MIT.
Afterwards {\tt Mathematica}~\cite{mathematica}, {\tt Maple}~\cite{maple} 
and others have been developed which are still in use nowadays. However, 
their field of application is limited to small and medium sized problems 
since it is not possible to work with very large intermediate expressions. 
On the other hand, there are quite a number of problems which produce 
intermediate expressions of the order of a few hundred giga bytes up to 
tera bytes to be manipulated by the CAS. The only CAS currently available 
in order to cope with such tasks is \form{}~\cite{Vermaseren:2000nd,
Kuipers:2012rf}.

\form{} is a program for the 
symbolic manipulation of algebraic expressions. It is specialized to handle 
very large algebraic expressions of billions of terms in an efficient and 
reliable way. That is why it is widely used, in particular in the framework 
of perturbative Quantum Field Theory, where often several thousands of 
Feynman diagrams have to be computed. However, the abilities of \form{} are 
also quite useful in other fields of science where the manipulation of huge 
expressions is necessary.

\form{} is constructed in such a way that the size of the expressions is 
not restricted by the main memory of the computer but only by the space 
available on hard disk. In addition its data representation is very dense 
when compared to other general purpose systems. Actually in modern 
applications in particle physics it happens quite often that the size of 
intermediate expressions for each Feynman diagram may become huge. As a 
consequence, even with \form{} such calculations require a CPU time of 
several years despite the steady advancement of the hardware and the 
continuous improvement of the algorithms.
Furthermore the resources as far as CPU speed, memory and disk 
space are concerned are often not sufficient.

One of the most efficient ways to increase the performance is based on 
parallelization which makes simultaneously available the 
resources of several computers and thereby significantly reduces the wall 
clock time. In fact, the project to obtain a parallel version of 
\form{} has been started at the end of the nineties. In the recent years 
\parform{}~\cite{Tentyukov:2004hz} and \tform~\cite{Tentyukov:2007mu} have 
become reliable tools which shall be described in this contribution.

There is a number of calculations performed within project A1 of the 
CRC/TR~9 where \parform{} and \tform{} were essential for the successful 
completion~\cite{Baikov:2002uw,Baikov:2002va,Baikov:2003zg,Baikov:2003gu,Baikov:2004tk,Chetyrkin:2005kn,Baikov:2005rw,Baikov:2006ch,Baikov:2008jh,Baikov:2009bg,Baikov:2010je,Baikov:2012er,Baikov:2012zm,Baikov:2012zn,Baikov:2014qja}.
In all these cases the single-core CPU time was estimated to several years. 
Parallelization could reduce the wall clock time to weeks and months at 
most.

As a further application we want to mention Ref.~\cite{Blumlein:2009cf} 
where \form{} was used to solved exceptionally large systems of equations 
to create mathematical tables for general use in mathematics and physics.

The calculation of three-loop helicity-dependent splitting functions in 
QCD~\cite{Vogt:2014pha,Moch:2014sna} also could only be completed thanks to \form{} 
because expressions of one tera byte or more were no exception and at one 
point more than 6 tera bytes of diskspace was needed for a single diagram.

Within the CRC/TR~9 two concepts for parallel versions of \form{} have 
been successfully developed and implemented: \parform{}, essentially 
based on MPI (message passing interface), and \tform{} which uses 
threads for the parallelization. Both programs run stable, show a good 
speedup and are complete in the sense that all programs written for the 
serial version of \form{} can now be used with \parform{} and \tform{}. 
In Sections~\ref{sec::parform} and~\ref{sec::tform} details to the 
parallel versions are provided.

In this project of the CRC/TR~9 also programs concerned with the reduction 
of families of Feynman integrals to a small set of basis elements (master 
integrals) and their numerical evaluation have been developed. These two 
topics are covered in two program packages, {\tt FIRE} and {\tt FIESTA}, 
which are discussed in Section~\ref{sec::other}

We continue this review in Section~\ref{sec::hist} with some historical 
remarks concerning the first steps towards parallelization of \form{} 
and describe in Section~\ref{sec::form} the basic features of \form{}.

%- }}}
%- {{{ Historical remarks:

\section{\label{sec::hist}Historical remarks}

The first initiatives of parallizing \form{} go back to early 1991, when 
version 1 of \form{} was made to run on a computer at the Fermi National
Accelerator Laboratory (FNAL) which was 
designed for lattice calculations and had 257 processors. Due to 
limitations in accessibility this project was discontinued, but the further 
development of \form{} took this experience into account.

The first systematic study of a parallel version of \form{} has been 
performed within the DFG-funded Research Unit ``Quantenfeldtheorie, 
Computeralgebra und Monte-Carlo Simulation'' which ran from 1996 to 2002 
and thus can be considered as a precursor to the CRC/TR~9. In 
Ref.~\cite{Fliegner:1999jq} a first parallel prototype of \form{} has 
been presented and results for several studies like the runtime for the 
parallel sorting on different architectures are shown.

One year later, in July 2000, the first ``working parallel \form{} 
prototype, \parform{}'', has been introduced in 
Ref.~\cite{Fliegner:2000uy}. It was based on the syntax of a preliminary 
version of \form{}~3 which at that time was not published yet. 
In~\cite{Fliegner:2000uy} the parallelization on clusters has been 
discussed based on the following hardware:
\begin{itemize}
\item 
  Digital workstation cluster (TTP Karlsruhe) running DEC UNIX 4.0D
  8 nodes with 600~MHz Alpha 21164A (EV56) processors and 512~MB RAM,
\item 
  PC cluster (TTP Karlsruhe) running Linux 2.2.13
  4 nodes with 500~MHz Intel Pentium III processors and 256~MB RAM,
\item
  IBM SP2 (Computing Center Karlsruhe) running AIX 4.2.1
  160 thin P2SC nodes with 120~MHz processors and 512~MB RAM (256 nodes in total).
\end{itemize}
Next to several feasibility studies also results for the speedup of a {\tt 
MINCER}~\cite{Larin:1991fz} job is shown. A reasonable speedup of 2.5 with 
four nodes on the PC cluster, a factor of 4.5 with eight nodes on the Alpha 
cluster and a factor of 6 with twelve nodes on the IBM SP2 has been 
reported. As a first physical application of \parform{} higher moments 
of deep inelastic structure functions at next-to-next-to-leading order of 
perturbative QCD have been computed in Ref.~\cite{Retey:2000nq}.

\begin{figure}[t]
  \begin{center}
    \includegraphics[width=\linewidth]{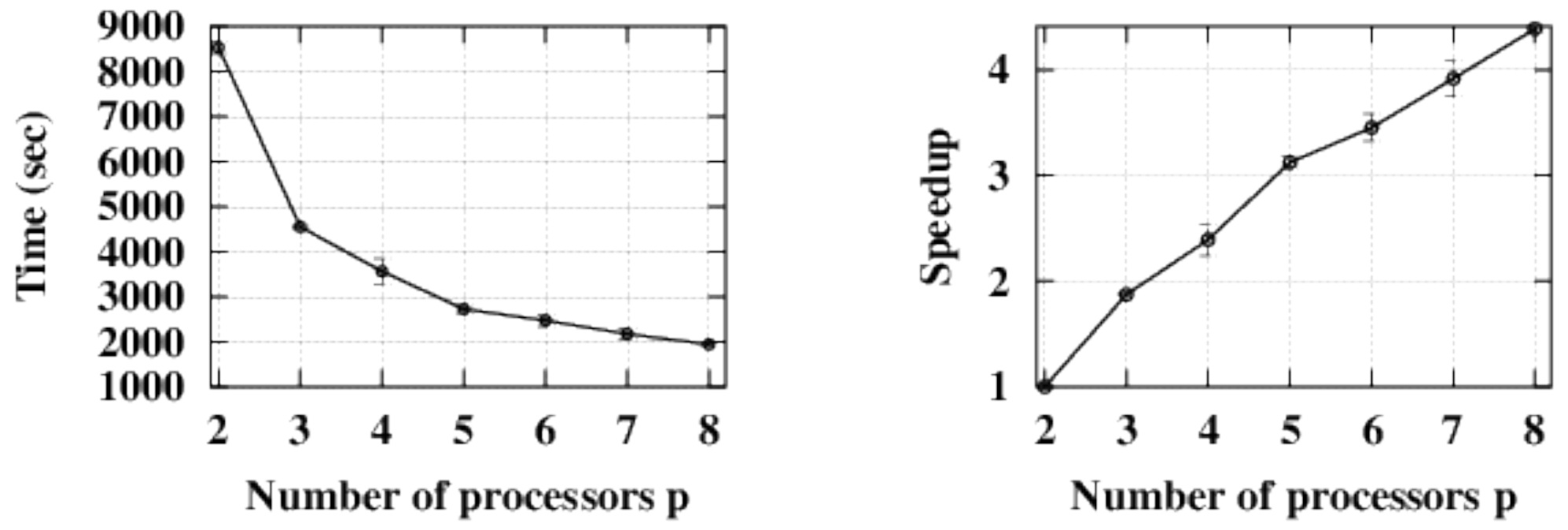}
    \caption{\label{fig::speedup_qcmsmp_2002}Speedup for the program {\tt BAICER} on
      Compaq-AlphaServer with 8 Alpha (EV67) processors with 700 MHz.}
  \end{center}
\end{figure}

At a later stage of the Research Unit \parform{} was further developed 
and one could run parallel \form{} jobs on symmetric multiprocessing 
(SMP) computers (not only on clusters). In 
Fig.~\ref{fig::speedup_qcmsmp_2002} the speedup is shown for the test 
program {\tt BAICER}, a \form{} program developed to compute massless 
four-loop two-point integrals within the project A1 of the CRC/TR~9, 
running on a
\begin{itemize}
\item
  Compaq-AlphaServer GS60e, 8 Alpha (EV67) processors (700 MHz).
\end{itemize}
A speedup of about 4.5 could be achieved using eight processors.

Two years after the start of the CRC/TR~9 a first version of \parform{}
operating on Cluster- and SMP-architectures was discussed in
Ref.~\cite{Tentyukov:2004hz}. It could run arbitrary \form{} programs in
parallel and was based on \form{}~3 version 3.1~\cite{Vermaseren:2000nd}. At
that time there were already a number of applications which would not have
been possible
without \parform{}~\cite{Retey:2000nq,Baikov:2002uw,Baikov:2002va,Baikov:2003gu}.

\begin{figure}[t]
  \begin{center}
    \includegraphics[width=\linewidth]{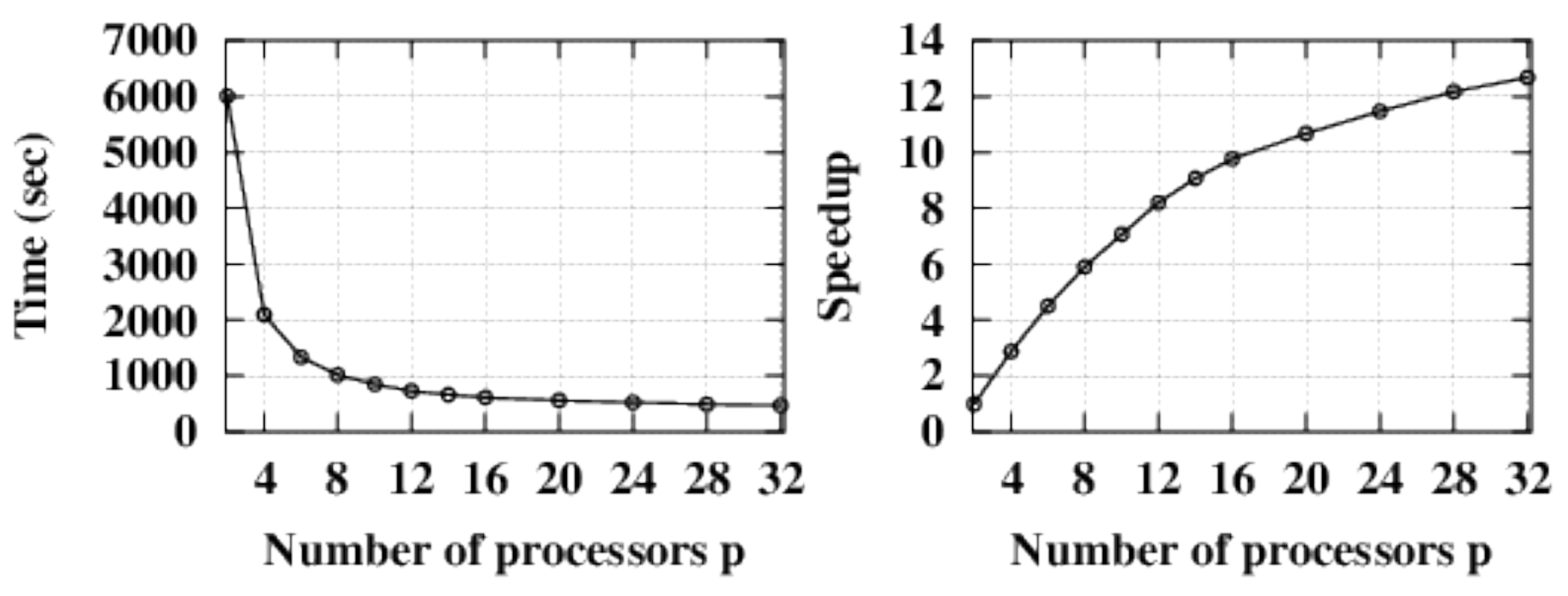}
    \caption{\label{fig::speedup_sgi_2004}Computing time and speedup for the
      test program BAICER on the SGI Altix 3700 server with 32x Itanium-2
      processors (1.3 GHz).}
  \end{center}
\end{figure}

For the calculations and for the development of \parform{} a 32-core
computer was available
\begin{itemize}
\item
  SGI Altix 3700 Server 32x 1.3 GHz/3~MB-SC Itanium-2 CPUs
  64 GB DDR/116 MHz mem, 2.4 TB SCSI hard disks.
\end{itemize}
The results for the test program {\tt BAICER} are shown in 
Fig.~\ref{fig::speedup_sgi_2004}. The speedup is almost linear up to twelve 
processors. Afterwards it flattens but is still considerable.  An achieved 
speedup of 12 means that a \form{} job that would need one year of 
computing time can be run as \parform{} job in about one month. This 
leads to a qualitatively new level, because it would practically be 
impossible to run jobs for years whereas months are feasible nowadays. 
Fig.~\ref{fig::speedup_sgi_2004} shows that with 16 processors a speedup of 
10 could be reached. This means that one can run on a 32-processor computer 
two jobs simultaneously, having the speedup of 10 for each of them.

In the paper~\cite{Tentyukov:2006ys} the functionality of \form{} and 
\parform{} was extended and facilities were introduced to communicate 
with external resources. This mechanism enables the user to include into 
the \form{} programs other pieces of software which are used as black 
box in order to take over certain tasks. As a typical example we want to 
mention is {\tt fermat}~\cite{fermat}, which can compute the greatest 
common divisor of multi-variable polynomials efficiently.

In February 2007 \tform{}~\cite{Tentyukov:2007mu} based on {\tt POSIX} 
threads has been released, a further major step in the development of 
parallel \form{} versions.
For later developments and further comparisons between \parform{} and \tform{}
we refer to the proceedings
contributions~\cite{Vermaseren:2006ag,Tentyukov:2008zz,Tentyukov:2010qf,Tentyukov:2006pr}
and to Sections~\ref{sec::parform} and~\ref{sec::tform}.

The more recent developments concern the release of
\form{}~4.0~\cite{Kuipers:2012rf} and the inclusion of tools to generate
optimized  
code~\cite{Kuipers:2013pba} which is used as input in {\tt FORTRAN} or {\tt 
C} programs for numerical integrations.

%- }}}
%- {{{ FORM:

\section{\label{sec::form}Sequential version of \form{}}

This article is not intended as an introduction to \form{} or even a  
reference manual. Nevertheless we want to describe the basic features which  
are important in the context of parallelization. 
 
\begin{figure}[t]
  \begin{center}
    \includegraphics[width=.45\textwidth]{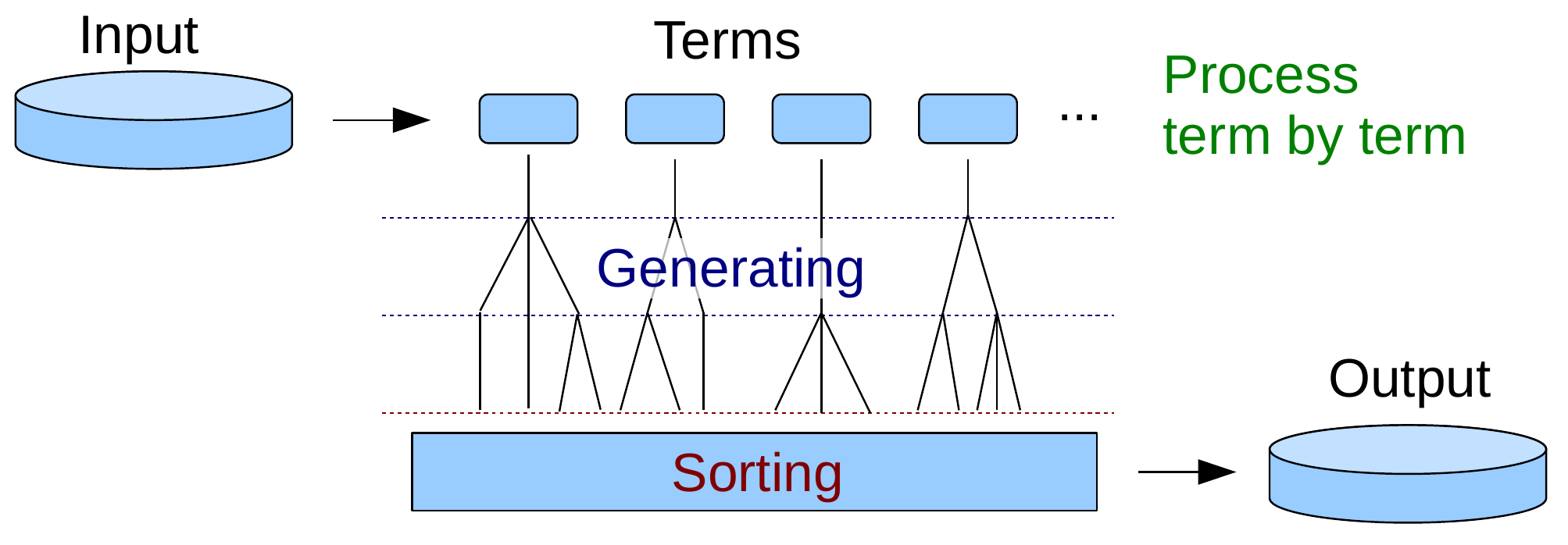}
    \caption{\label{fig::form_stream}
      Graphical representation of the processing of an input expression in \form{}.
    }
  \end{center}
\end{figure}

A \form{} program is in general divided into so-called modules which are 
terminated by a ``dot''-instruction. During the execution of the program, 
which is only possible in batch-mode, each module is processed 
separately one after the other which essentially occurs in three steps
\begin{itemize}
\item{Compilation:} The input is translated into an internal 
  representation.
\item{Generating:} For each term of the input expressions the statements of 
  the module are executed. This in general generates a lot of terms.
\item{Sorting:} All the output terms that have been generated are sorted 
  and equivalent terms are summed up.
\end{itemize}
This is illustrated in Fig.~\ref{fig::form_stream}.

The fundamental objects which are manipulated by \form{} commands are 
expressions which are viewed as sums of individual terms (see also 
Fig.~\ref{fig::form_stream}). Next to a sophisticated pattern matcher, it 
is the strength of \form{} that only local operations on single terms are 
allowed, like replacing parts of a term by some other expressions. 
Non-local operations like replacing a sum of two terms are not allowed. For 
example, the command \verb|identify| (short: \verb|id|) identifies the 
left-hand side with the right-hand side and can be used as
\begin{verbatim}
  id a = b + c;
\end{verbatim}
On the other hand, the usage
\begin{verbatim}
  id a + b = c;
\end{verbatim}
would lead to an error message.

Non-local operations are allowed only implicitly, e.g., in the sorting 
procedure at the end of the modules, where equivalent terms are combined. 
At first sight this seems to be a strong limitation for the formulation of 
general and efficient algorithms. It is usually possible to get around this 
limitation by designing algorithms in clever and non-standard ways.

Due to the locality of the operations it is possible to handle expressions 
as ``streams'' of terms that can be read sequentially from the memory or a 
file and processed independently. This enables \form{} to deal with 
expressions that are larger than the available main memory.

\begin{figure}[t]
  \begin{center}
    \includegraphics[width=.45\textwidth]{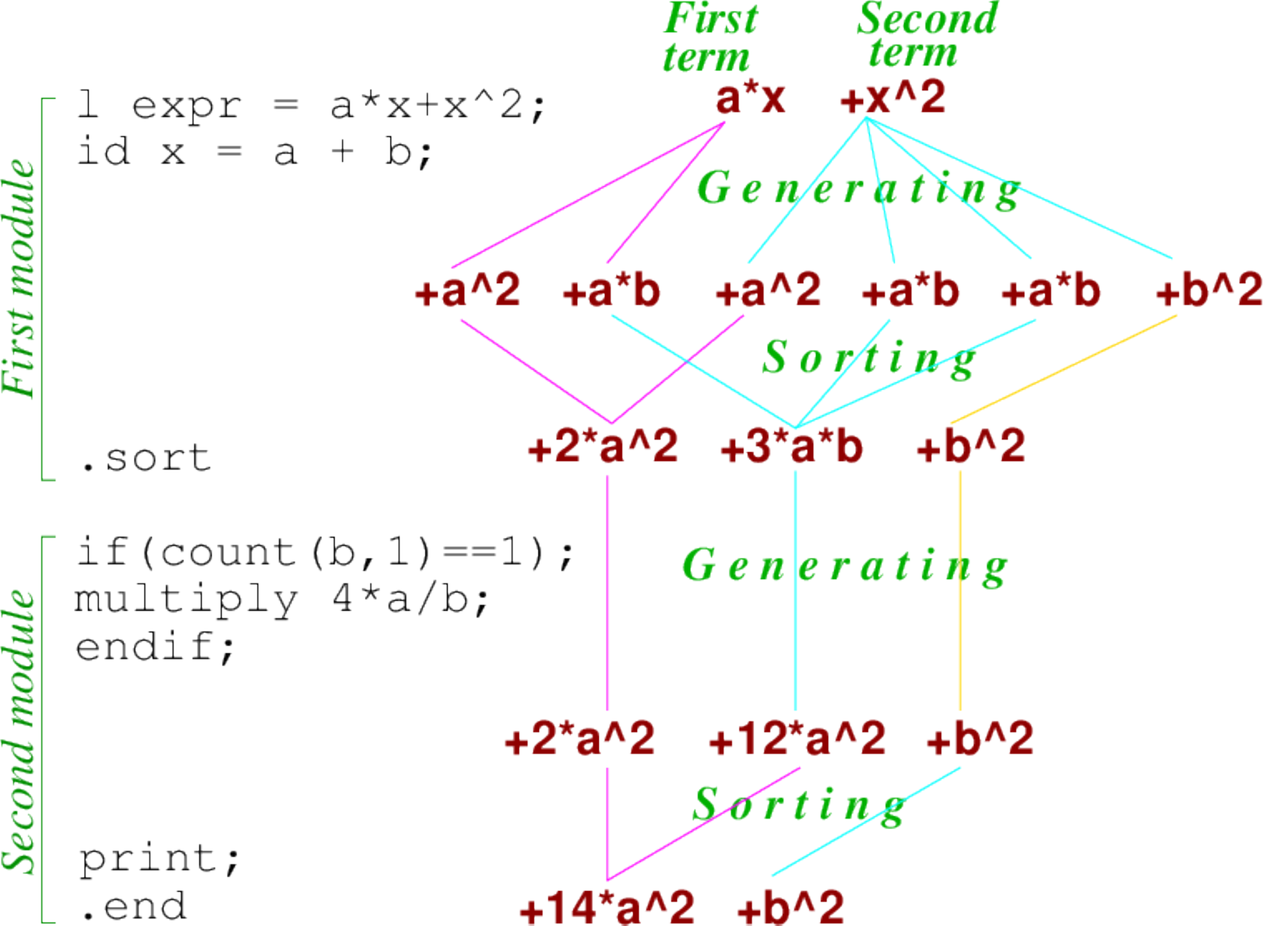}
    \caption{\label{fig::form_example}
      Example for the generation and sorting of data in \form{}.
    }
  \end{center}
\end{figure}

An example illustrating the principle operating mode of a \form{} program 
is shown in Fig.~\ref{fig::form_example}. It corresponds to the simple 
program
\begin{verbatim}
  l expr = a*x + x^2;
  id x = a + b;
  .sort
  if (count(b,1)==1);
    multiply 4*a/b;
  endif;
  print;
  .end
\end{verbatim}

%- }}}
%- {{{ ParFORM:

\section{\label{sec::parform}\parform{}}

\subsection{The concept of \parform{}}

\begin{figure}[t]
  \begin{center}
    \includegraphics[width=.45\textwidth]{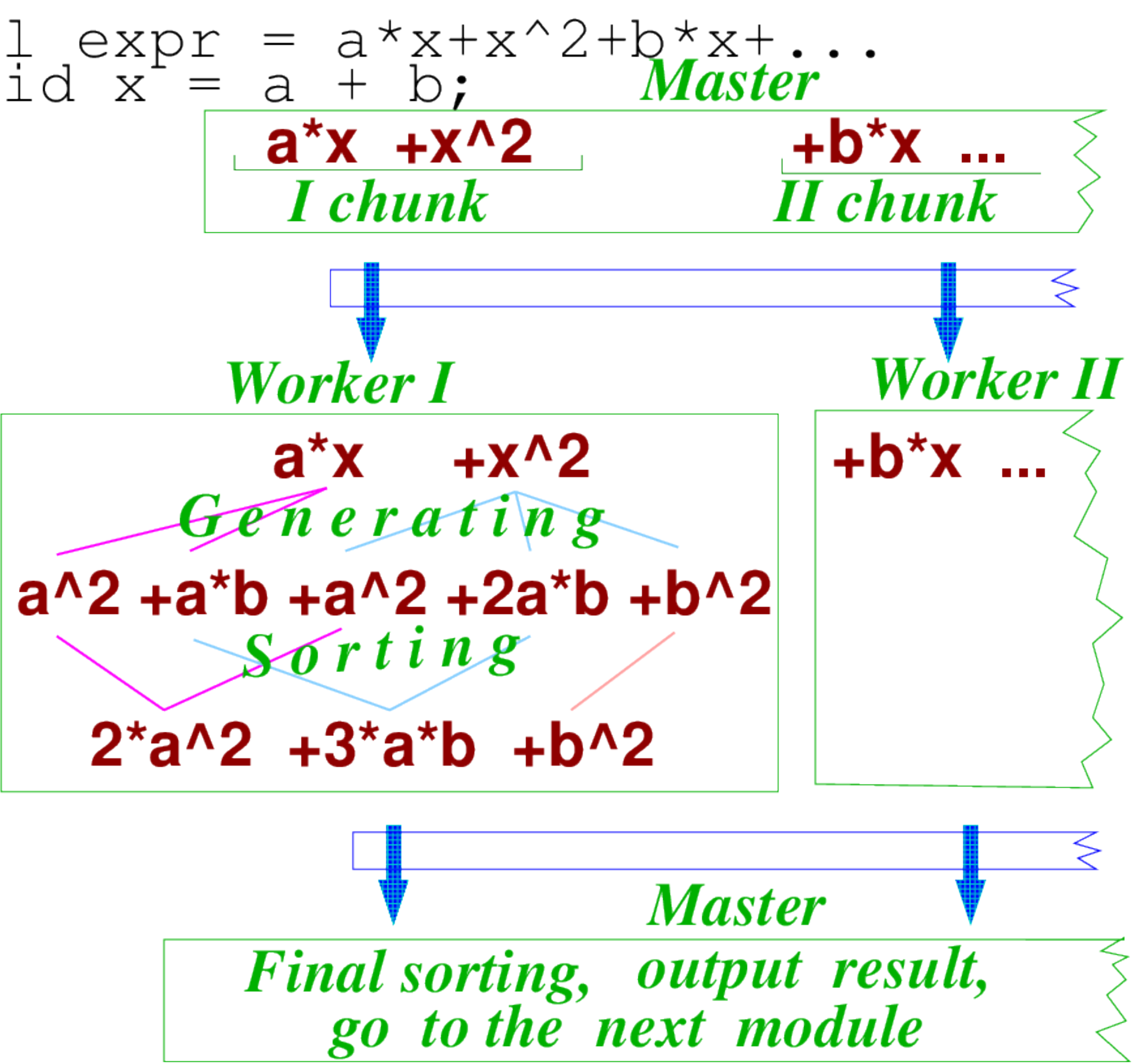}
    \caption{
      \label{fig::parexample}
      General conception of \parform{}.
    }
  \end{center}
\end{figure}

As mentioned above, the locality principle enables \form{} on the one hand 
to deal with expressions that are larger than the available main memory, on 
the other hand it also allows for parallelization. The concept implemented 
in \parform{} is straightforward and indicated in 
Fig.~\ref{fig::parexample}: in a first step the master process splits the 
expression into pieces, so-called chunks. Each chunk is sent to one of the 
workers where an independent \form{} process runs, i.e. the module to be 
executed is compiled, the terms are generated, sorted and sent back to the 
master. Once all worker processes have finished their jobs the master 
performs the final sorting.

\begin{figure}[t]
  \begin{center}
    \includegraphics[width=.45\textwidth]{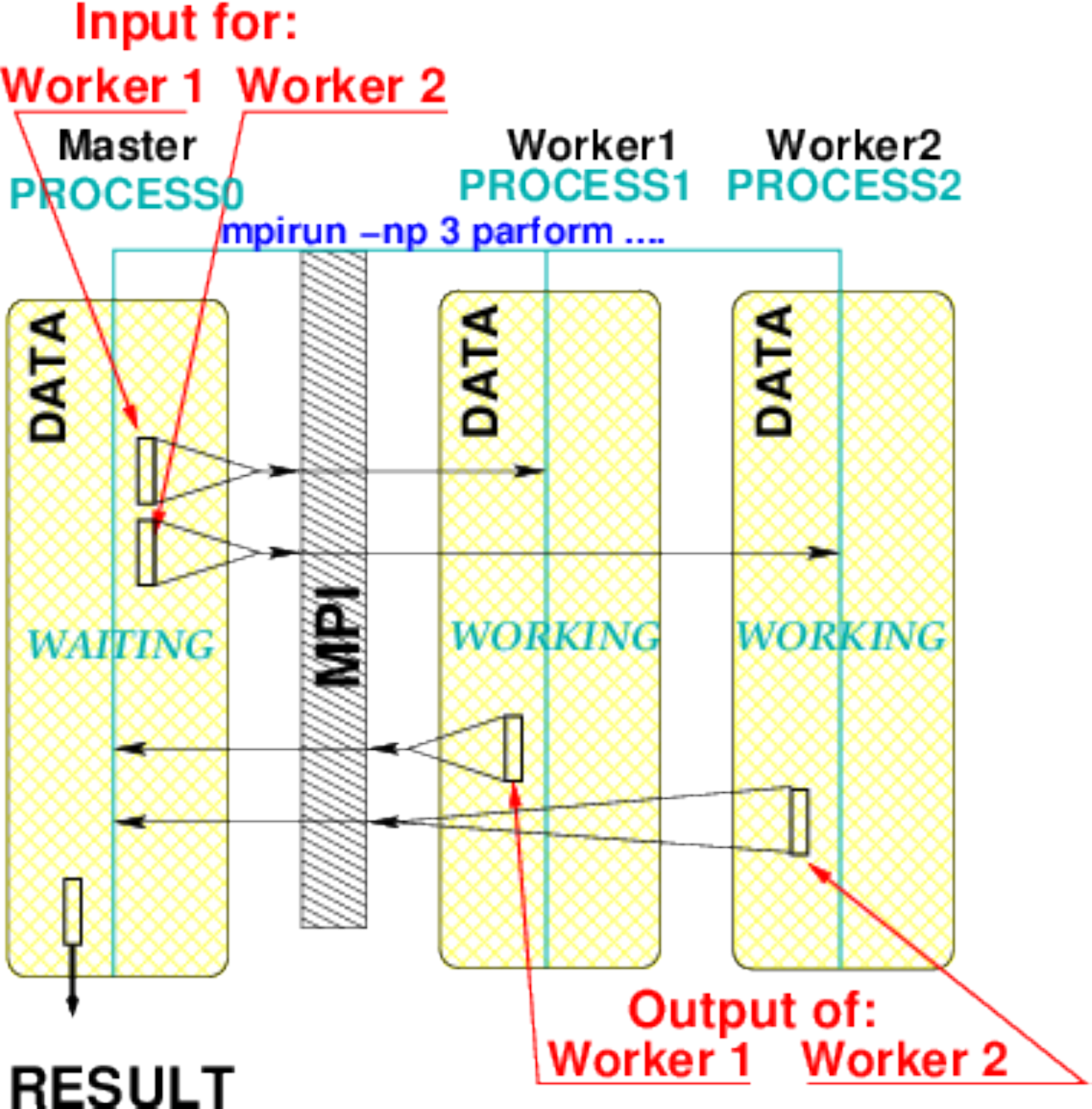}
    \caption{
      \label{fig::parform_mpi}
      Visualization of the mode of operation of \parform{} based on MPI.
    }
  \end{center}
\end{figure}

The communication between master and workers is based on the message 
passing interface (MPI) standard~\cite{MPI} which provides a library for 
the data transfer between processes. Message passing permits to parallelize 
\form{} on computer architectures both with shared memory, i.e. 
SMP computers and on computer clusters. The way the 
master communicates with the workers is sketched in 
Fig.~\ref{fig::parform_mpi}.

It is worth mentioning that the parallelization does not require any 
additional efforts from the user. It is possible to run the programs 
written for the sequential version using \parform{} and adding a 
specification concerning the number of processors. It is clear that 
different codes show a different performance and efficiency in the parallel 
version. In particular, modules in which the outcome depends on the order 
in which the terms are processed cannot be parallelized and are executed in 
sequential mode. This concerns mostly the use of the dollar variables which 
were introduced in version 3. In the case that \form{} would switch to 
sequential mode, while actually this is not needed, the user can add an 
extra statement to overrule such a decision and tell \form{} how to deal 
with the `dubious' case.

\subsection{\label{sub::numa}\parform{} on a NUMA architecture}

\begin{figure}[t]
  \begin{center}
    \includegraphics[width=.45\textwidth]{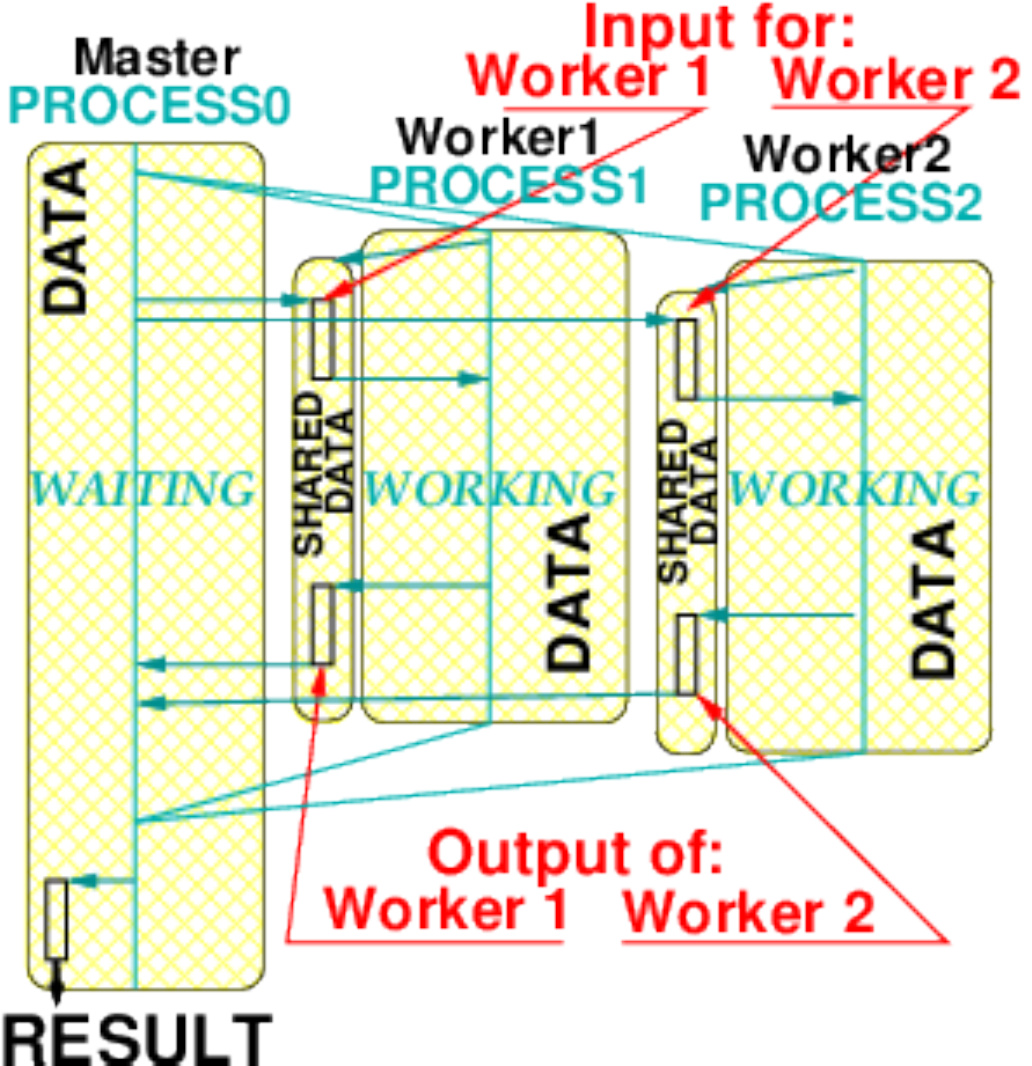}
    \caption{
      \label{fig::mpi_numa}
      Mode of operation implemented into \parform{} for a NUMA architecture.
    }
  \end{center}
\end{figure}

The SGI Altix computer is realized with a so-called NUMA architecture where 
NUMA stands for non-uniform memory access. This means that the individual 
processors have a faster access to some parts of the main memory than to 
others. A specialized version of \parform{} has been developed which 
exploits the feature and, at the same time, does not use MPI and the 
overhead connected to it. The corresponding scheme of operation is 
illustrated in Fig.~\ref{fig::mpi_numa}.

\begin{figure}[t]
  \begin{center}
    \includegraphics[width=.45\textwidth]{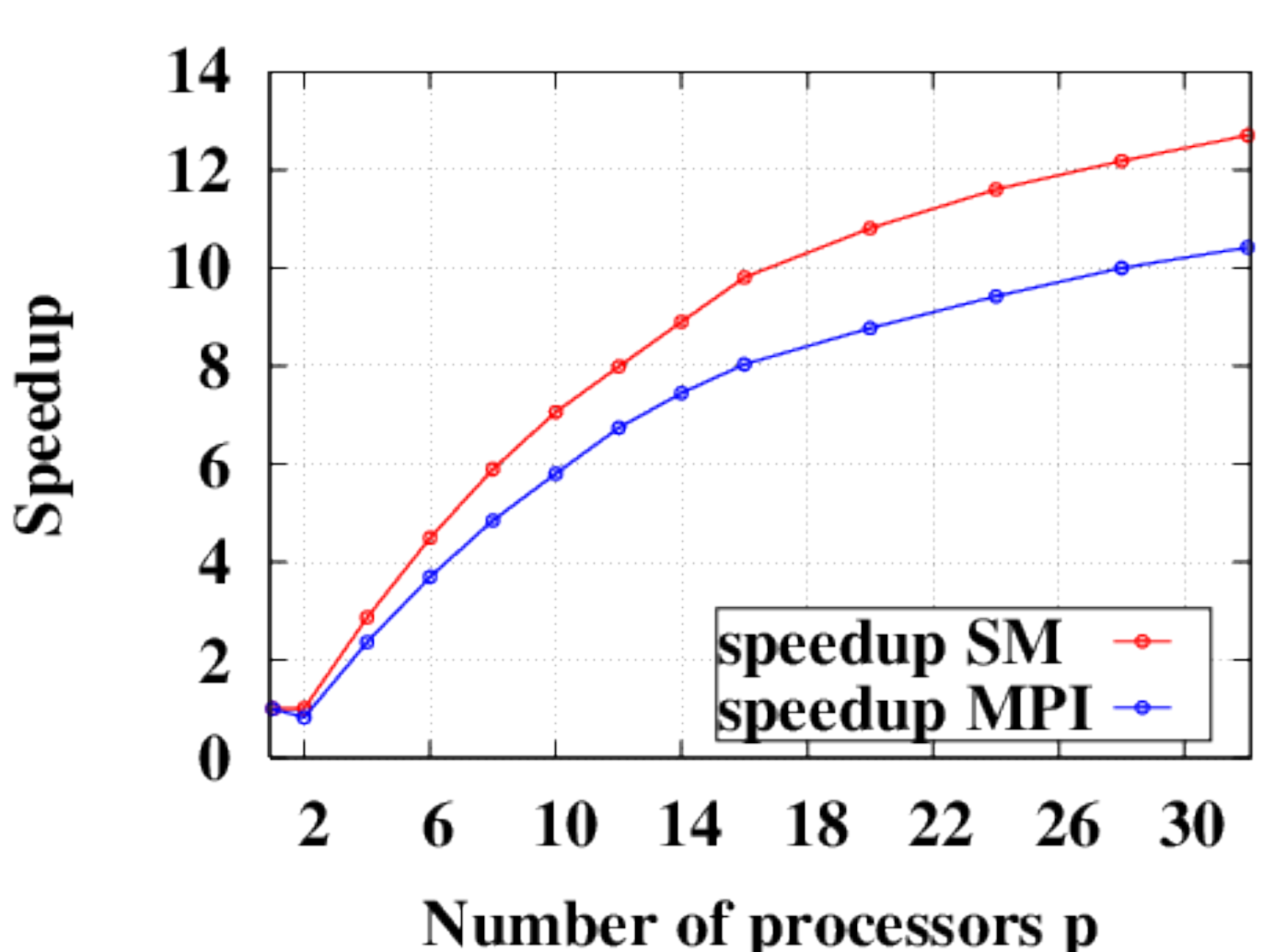}
    \caption{
      \label{fig::mpi_numa_speedup}
      Runtime and 
      speedup for the test program {\tt BAICER} running on a
      SGI Altix 3700 server with 32 Itanium-2 processors (1.3 GHz).
      The lower curve corresponds to the MPI version and the upper 
      one to the shared memory version of \parform{}.
    }
  \end{center}
\end{figure}

Using the specialized version of \parform{} in connection with the 
32-core SGI Altix a considerable improvement of the speedup could be 
obtained, as can be seen in Fig.~\ref{fig::mpi_numa_speedup}. In fact, for 
16 processors the speedup improved from 8 to 10, for 32 processors from 10 
to 13 (see also the discusion in the next subsection).

\subsection{\parform{} on clusters and multi-core nodes}
 
At present, there are a number of calculations of physical quantities which
would not have been possible without the gain in performance and speedup
provided by \parform{} (see, e.g.,
Refs.~\cite{Retey:2000nq,Baikov:2002va}). Most of the applications are
connected to the evaluation of four-loop Feynman integrals which occur in the
context of perturbative quantum field theory. In particular, there are
algorithms which transform the mathematical complexity of the original problem
to the need of simple manipulations of rather large polynomial expressions
which have billions or even more terms. Manipulations of this type constitute
the basis of the speedup curves which are discussed in the following.

The results for the test program running on a SGI Altix 3700 server with 32 
Itanium-2 processors are shown in Fig.~\ref{fig::mpi_numa_speedup} where 
both the runtime and the speedup (as compared to the sequential version) is 
shown as a function of the number of processors, $p$, involved in the 
calculation. The almost horizontal line between $p=1$ and $p=2$ is due to 
the fact that for $p=2$ one of the processors takes over the role of the 
master and the other one of the worker. Thus a real reduction of the CPU 
time only starts from $p=3$. It is interesting to note that the speedup is 
almost linear up to twelve processors. Furthermore, for 16 processors the 
program is faster by an order of magnitude. As a consequence instead of 
years one only has to wait a few months in order to obtain the results of a 
calculation. This provides the possibility to consider qualitatively new 
kinds of problems, since in practice it is impossible to run a job for 
years whereas a few months are feasible nowadays. Beyond $p=16$ the curve 
becomes more flat, however, the speedup is still considerable up to 32 
processors.

\begin{figure}[t]
  \begin{center}
    \includegraphics[width=.4\textwidth]{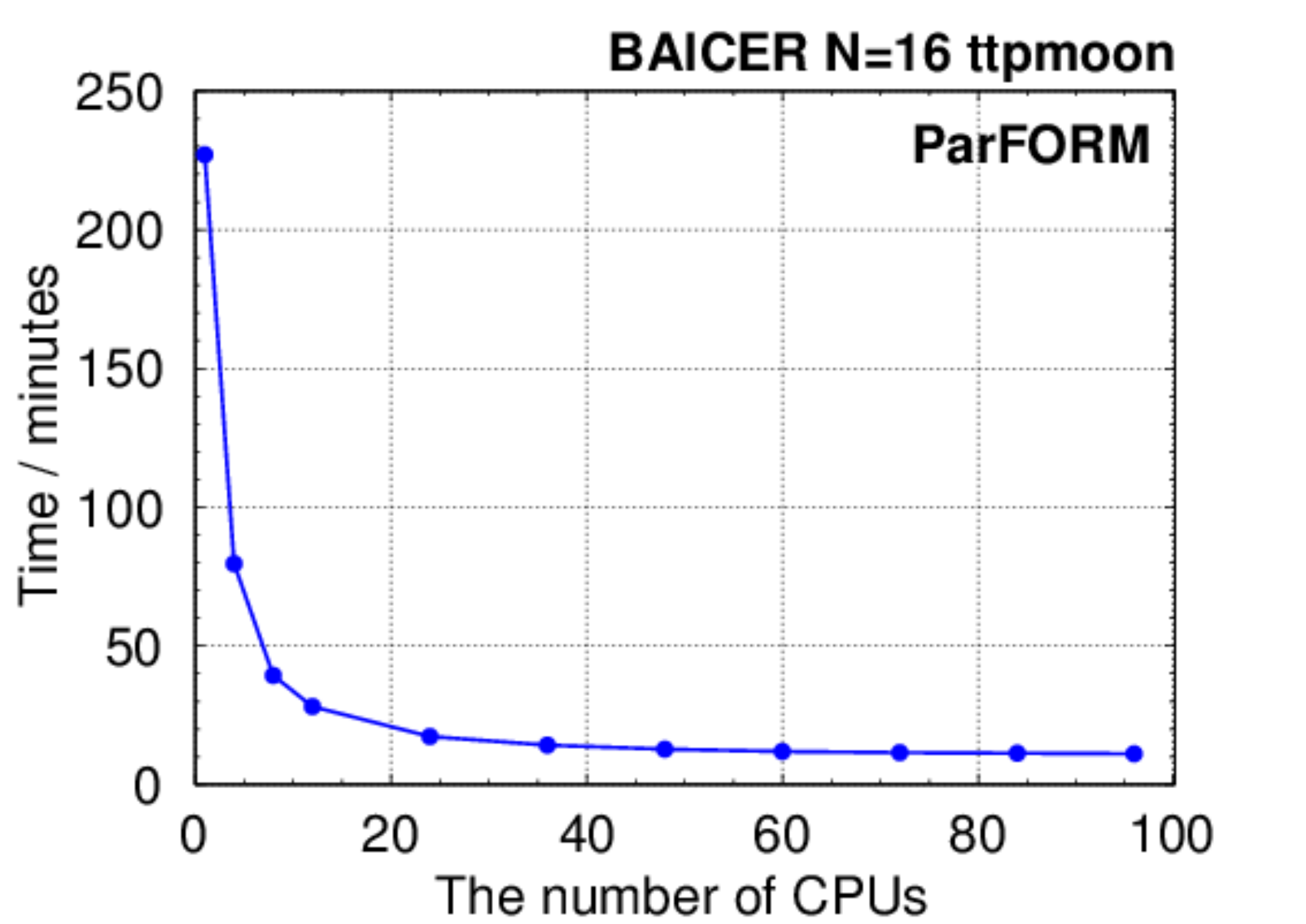}
    \\
    \includegraphics[width=.4\textwidth]{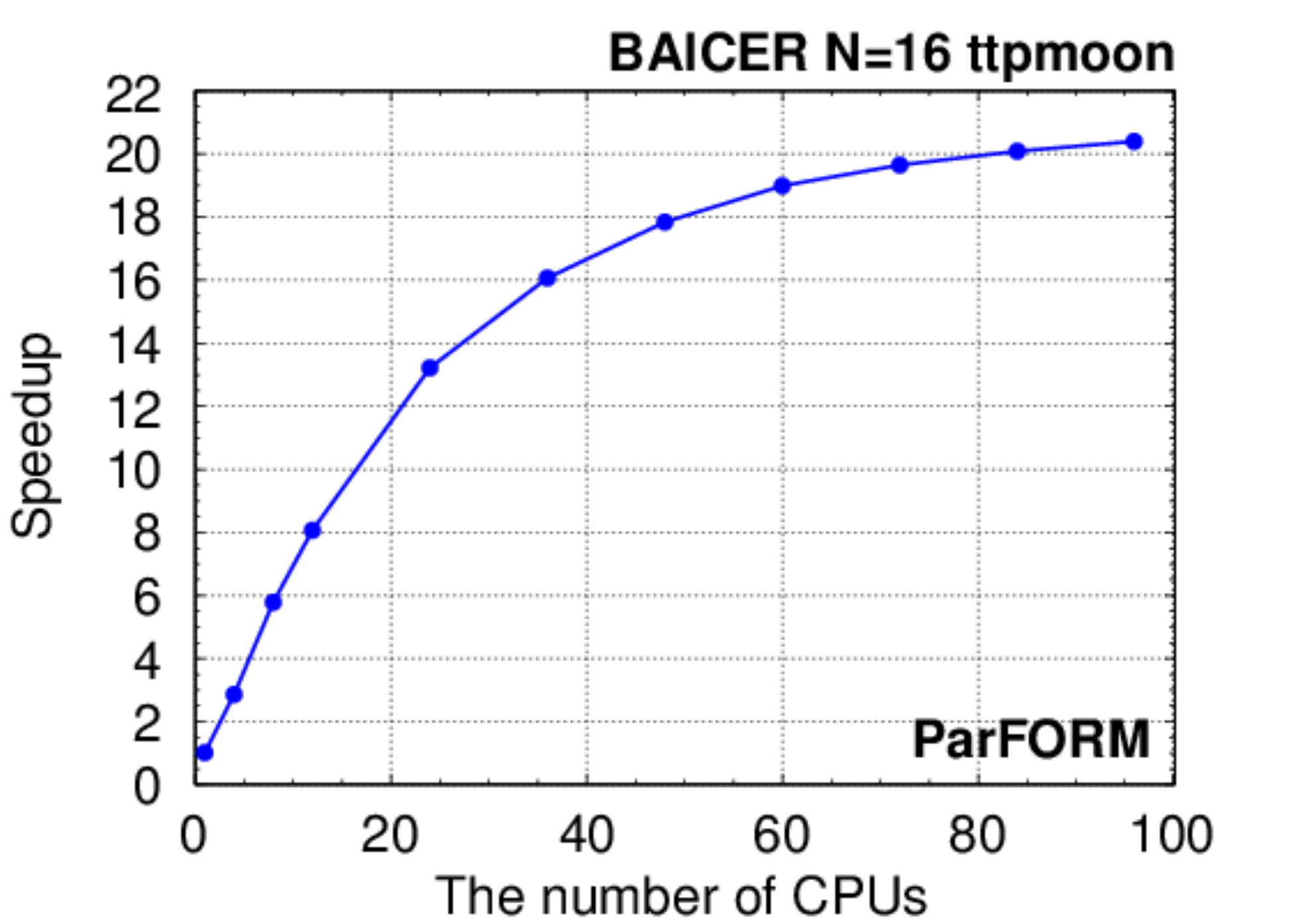}
    \caption{
      \label{fig::speedup_parform}
      Timing and speedup plot for the \parform{} benchmark job
      {\tt BAICER} running of {\tt ttpmoon}.
    }
  \end{center}
\end{figure}

The latest speedup plot for (the MPI version of) \parform{} is shown in
Fig.~\ref{fig::speedup_parform} where {\tt BAICER} is running on the cluster
{\tt ttpmoon} which has the following configuration:
\begin{itemize}
\item
  Computer cluster (TTP Karlsruhe) running Linux,
  8 nodes with
  2 Hexa-Core Intel Xeon X5675 (3.07 GHz),
  96 GB RAM,
  and 3.6 TB local hard disk (Raid 0 with 6 stripes),
  interconnected by QDR InfiniBand.
\end{itemize}
The top plot shows the used time in minutes as a function of the involved CPUs
(including the master) and on the bottom the speedup as compared to the serial
version is plotted.\footnote{Note, that there is no data point for two CPUs;
  otherwise one would observe a flat behaviour between one and two CPUs and
  only then the curve starts to raise.}  It is interesting to note that a
speedup of about 10 is reached in case 16 CPUs are used, a value obtained in
Fig.~\ref{fig::mpi_numa_speedup} for the shared memory version which avoids
the use of MPI, cf. Subsection~\ref{sub::numa}. For higher number of CPUs the
curve flattens but nevertheless reaches a speeup above 20 for 96 CPUs.

\subsection{\parform{} on ``low-level''  clusters}

\begin{figure}[t]
  \begin{center}
    \includegraphics[width=.4\textwidth]{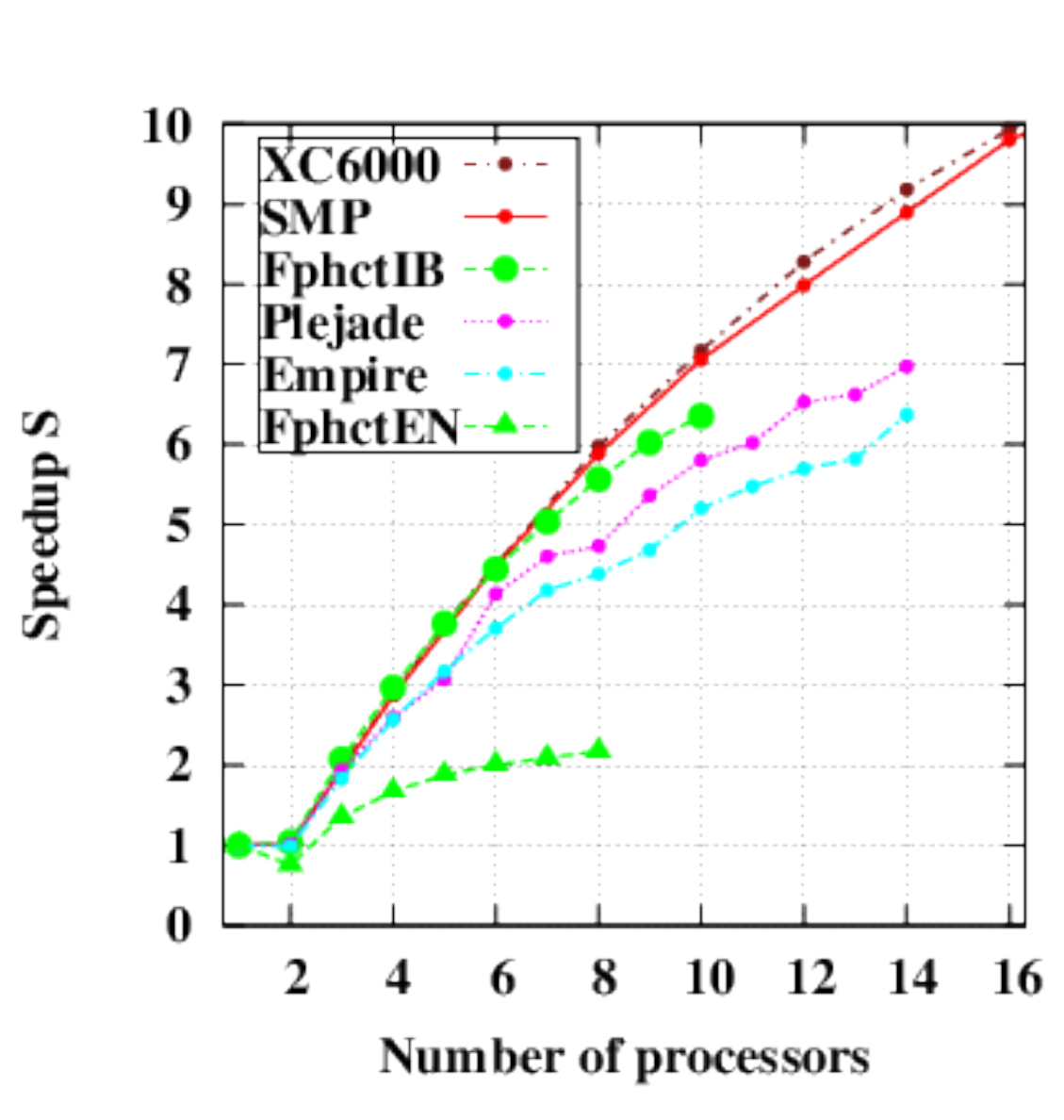}
    \caption{
      \label{fig::clusters}
      The speedup for the test program on different clusters in comparison
      to the SMP computer (cf. Fig.~\ref{fig::mpi_numa_speedup}).
    }
  \end{center}
\end{figure}

\parform{} has been successfully installed on several clusters. In 
Fig.~\ref{fig::clusters} the corresponding speedup curves are shown and 
compared to the curve from Fig.~\ref{fig::mpi_numa_speedup} obtained on the 
SMP computer. The cluster {\tt XC6000} is a Hewlett Packard Itanium-2 QsNet 
interconnected cluster.  This is the only tested cluster which demonstrates 
a better behaviour than the SMP computer, however, it is also significantly 
more expensive. {\tt Fphctl} is a cluster consisting of 32-bit Xeon nodes. 
This cluster has been tested both with an Infiniband ({\tt FphctlIB}) and a 
simple Fast Ethernet ({\tt FphctlEN}) interconnection. Whereas the latter 
is not of interest in practice the former shows a quite reasonable 
behaviour following closely the SMP curve for a smaller number of 
processors. {\tt Plejade} and {\tt Empire} are both dual Opteron clusters. 
However, {\tt Plejade} is interconnected using InfiniBand whereas {\tt 
Empire} uses Gigabit Ethernet. Both clusters show a reasonable behaviour 
leading to a speedup of about six for ten processors.

We want to mention that the SMP curves shown in Fig.~\ref{fig::clusters} 
are based on the shared-memory model mentioned above. On the other hand, 
for the clusters one has to rely on the MPI library which for our 
applications has a significant overhead.

%- }}}
%- {{{ TFORM:

\section{\label{sec::tform}\tform{}}

In the last decade multi-core processing has become a key technology in the 
computing industry as system performance improvement through increasing 
clock rates of single-core processors is hindered by physical limits. From 
laptops to supercomputers multi-core processing is prevalently used and the 
modern operating systems allow one to easily use them as SMP computers. 
Although \parform{} works on such SMP computers, interprocess 
communications among the master and the workers via MPI can have a 
significant overhead when gigantic expressions are transfered.

This overhead problem can be overcome on SMP architectures with the help of
another model for the communication. In this approach the master explicitly
allocates shared memory buffers which can be accessed both by the master and
the workers. In these memory segments the master prepares the chunks for the
workers, they are doing their job and the master collects the results again
from the shared buffers. Thus, copying huge amounts of data is not necessary
any more. The use of the shared-memory model on SMP machines led to an
increase in the speedup of 20-25\%
(cf. Fig.~\ref{fig::mpi_numa_speedup}). This concept is taken even further in
\tform{}~\cite{Tentyukov:2007mu}, a multithreaded version of \form{}.

\begin{figure}[t]
  \begin{center}
    \includegraphics[width=.45\textwidth]{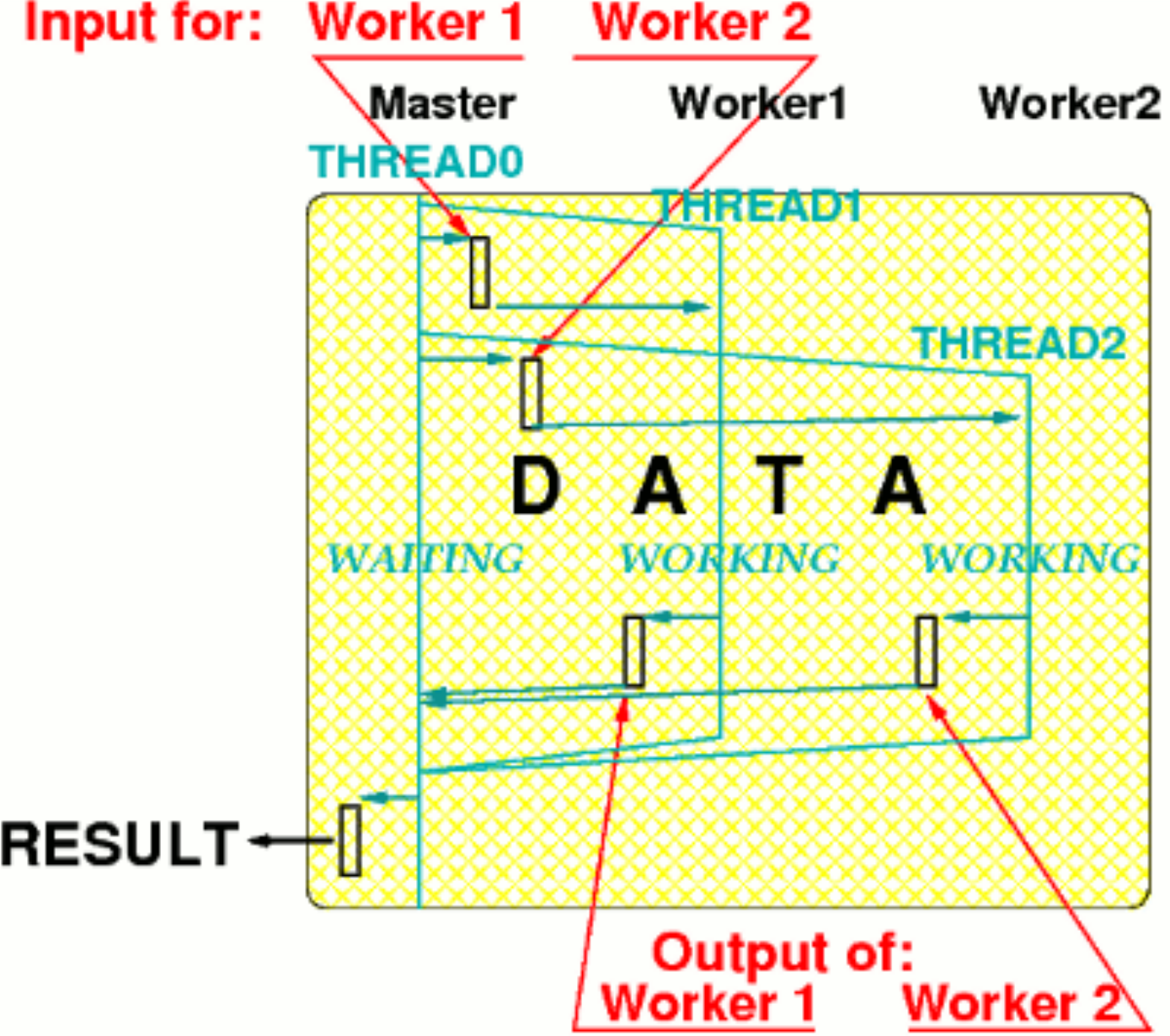}
    \caption{
      \label{fig::tform_thr}
      Mode of operation for \tform{}.
    }
  \end{center}
\end{figure}

In \tform{} the implementation uses the POSIX threads library, which is 
available on all modern UNIX systems and therefore portable. The way the 
master communicates with the workers is sketched in 
Fig.~\ref{fig::tform_thr}. \tform{} starts with one master thread and $N$ 
worker threads in a so-called thread pool.  The workers sleep until the 
master assigns tasks, and hence do not spend any CPU time.  When the master 
has some task to be distributed over the workers, the master wakes up one 
of the sleeping workers and assigns the task to it.  Terms in expressions, 
grouped as chunks for reducing the overheads, are distributed in this way.  
After distributing all terms, the master waits for all the workers to 
finish the tasks, and then the master merges the results of the workers in 
a final sorting operation. The data transfer among the threads is done via 
the shared memory buffers and by using memory locks for synchronization 
between the master and the workers (see Fig.~\ref{fig::tform_thr}).

Due to the model for the communications, some features improving the 
performance are relatively easy to implement in \tform{}, whereas their 
implementations are difficult in \parform{}. One of them is a load 
balancing system. If there is a single worker that is assigned terms 
requiring much CPU time, for the final sorting the master may have to wait 
for this worker even after the other workers finish their tasks and become 
idle. To avoid such inefficiency, after distributing all terms to be 
processed, the master looks for idle workers. If such workers are found, 
terms are stolen back from the chuncks of workers that are still busy and 
redistributed over idle workers. Experiments with an even more fine-grained 
load balancing were unsuccessful, because they resulted in too much 
overhead.

\begin{figure}[t]
  \begin{center}
    \includegraphics[width=.45\textwidth]{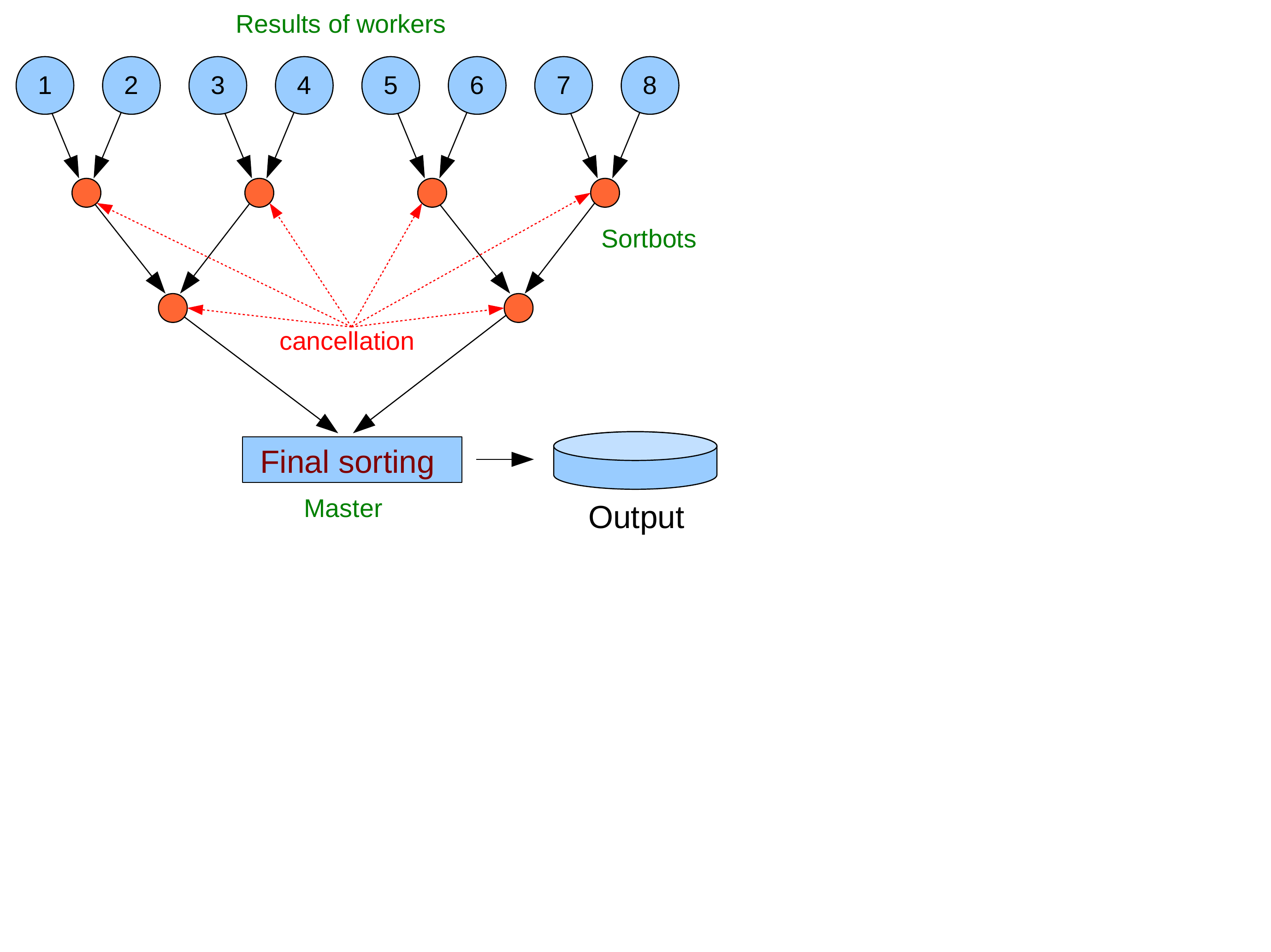}
    \caption{
      \label{fig::sortbots}
      Illustration of the mode of operation of sortbots.
    }
  \end{center}
\end{figure}

Another feature in \tform{} concerns the parallel sorting. In the final 
sorting, \tform{} used to adopt the simple model in which the master merges 
the outputs from all the workers simultaneously. Therefore it often happens 
that the master is busy while the workers are waiting for the master to 
accept their next chunks of the results. It becomes a bottleneck, 
especially when the number of the workers is large. To alleviate this 
bottleneck, an improved model of the final sorting has been implemented in 
\tform{}. In this model, each two workers send their results to a special 
worker thread, called a sortbot, which merges the results. Then each two 
sortbots send their results to another sortbot. This continues until the 
last two sortbots send their results to the master, which merges the final 
two results and writes the result to disk. This is illustrated in
Fig.~\ref{fig::sortbots}. 
Because also this method still involves much waiting, a run with $N$ 
workers will rarely use more than the CPU time provided by $N$ cores, even 
when the computer has many more cores. The total wall clock execution time 
improves measurably by this method, although it does go at the cost of 
extra memory needed for the buffers of the sortbots.

\begin{figure}[t]
  \begin{center}
    \includegraphics[width=.45\textwidth]{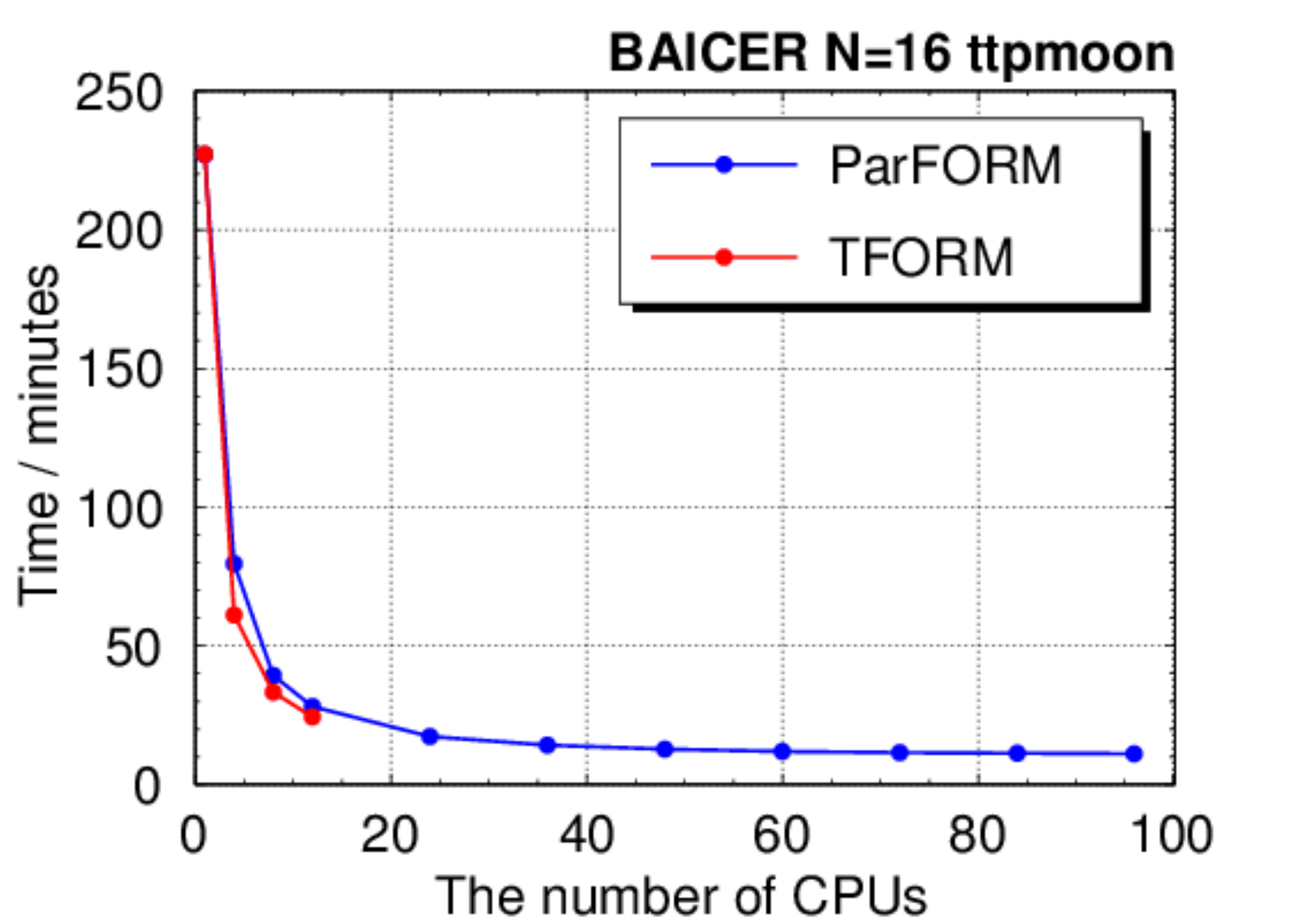}
    \\
    \includegraphics[width=.45\textwidth]{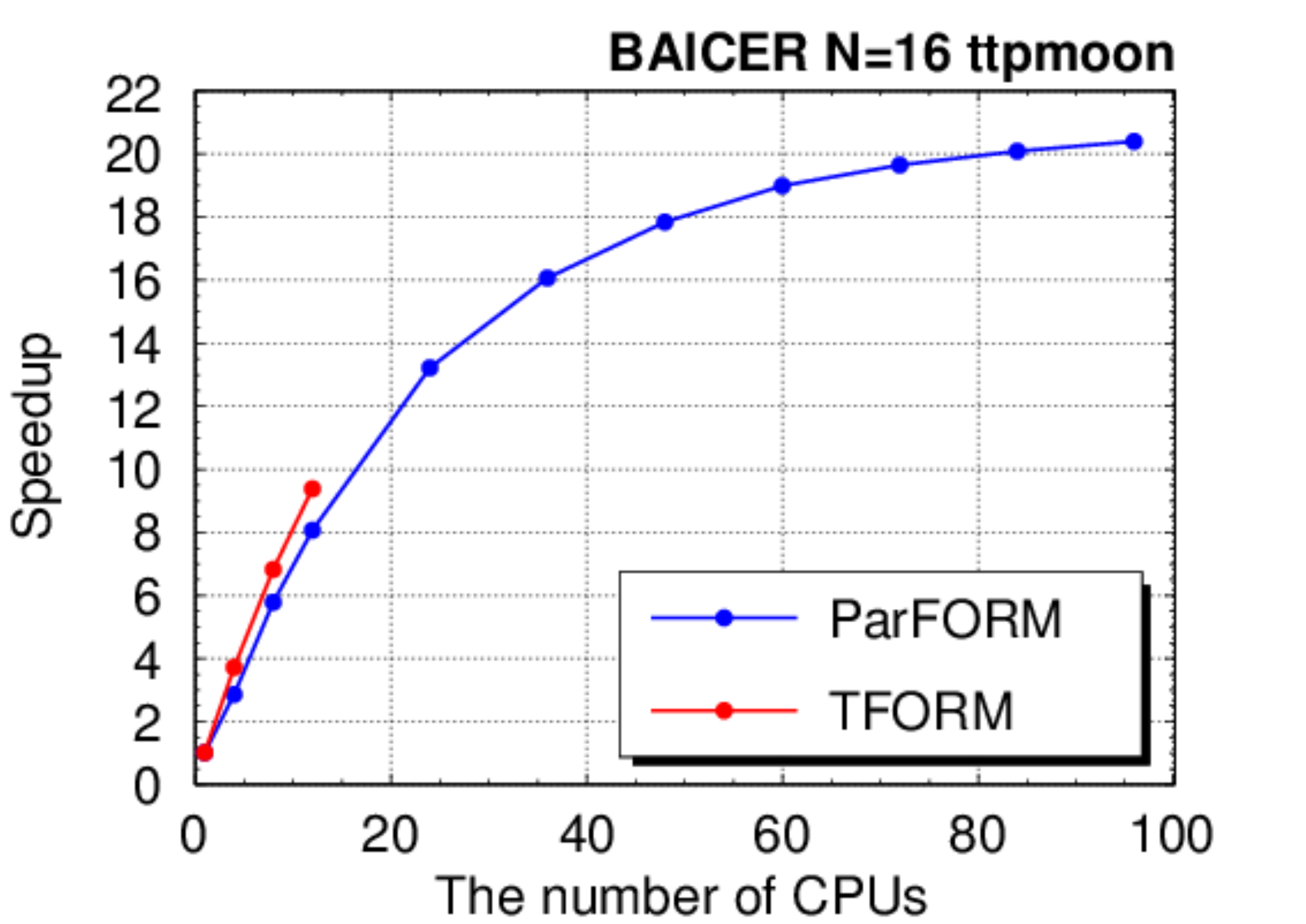}
    \caption{
      \label{fig::speedup_tform_parform}
    Comparison of timing and speedup for \parform{} and \tform{} running on
    {\tt ttpmoon}.}
  \end{center}
\end{figure}

Fig.~\ref{fig::speedup_tform_parform} shows up-to-date timing and speedup
plots for \parform{} and \tform{} running on {\tt
  ttpmoon}.\footnote{The \parform{} curves are already shown in
  Fig.~\ref{fig::speedup_parform}.} Note that the cluster {\tt ttpmoon}
consists of 12-core nodes which explains the end point of the \tform{} curves
where a speedup better than 9 is reached. \parform{} reaches for 12 CPUs,
which means 1 master and 11 workers, a speedup of~8.

%- }}}
%- {{{ Further developments:

\section{\label{sec::other}Further developments within CRC/TR~9}

\subsection{Reduction to master integrals with {\tt FIRE}}

Nowadays the vast majority of calculations of higher order quantum corrections
involve a huge number (sometimes exceeding several millions) of different
contributing integrals. The standard way to reduce their number to a
manageable amount is based on the so-called ``Laporta algorithm'' which is
described in Ref.~\cite{Laporta:2001dd}. There are many different
implementations of this algorithm, some of them are publicly available like
{\tt AIR}~\cite{Anastasiou:2004vj} or {\tt
  Reduze}~\cite{Studerus:2009ye,vonManteuffel:2012np}, others are private like
{\tt crusher}~\cite{crusher} which has been developed in the context of
project~A1 of the CRC/TR~9.  Within project~A2 the program {\tt
  FIRE}~\cite{Smirnov:2008iw,Smirnov:2013dia,Smirnov:2014hma,fire} has been
developed.
 
{\tt FIRE} stands for Feynman Integral REduction and implements a special
version of the Gauss elimination method to solve the system of linear equations,
which is generated by the application of the integration-by-parts
relations~\cite{Chetyrkin:1981qh}, for the master integrals. It uses several
external programs like {\tt Snappy}~\cite{snappy} for data compression,
{\tt KyotoCabinet}~\cite{kyotocabinet} as database to store data on disk, {\tt
  Fermat}~\cite{fermat} for algebraic simplifications, and {\tt
  LiteRed}~\cite{litered} to retrieve additional rules among integrals.

The operation of {\tt FIRE} is divided into two parts: in a first step the
input for the reduction step is prepared within {\tt Mathematica}. This
includes the generation of all integration-by-parts relations, the generation
of symmetry relations, the identification of the sectors of indices where
integrals vanish, i.e. the so-called boundary conditions, and the preparation
of a list of integrals which shall be reduced.  The second step is
significantly more time consuming. In the latest version, {\tt
  FIRE5}~\cite{Smirnov:2014hma}, this part is written in {\tt C++}. Here the
systematic reduction to master integrals is performed.  The output is a table
for the list of integrals provided in part one.

\begin{figure}[t]
  \begin{center}
    \includegraphics[width=.3\textwidth]{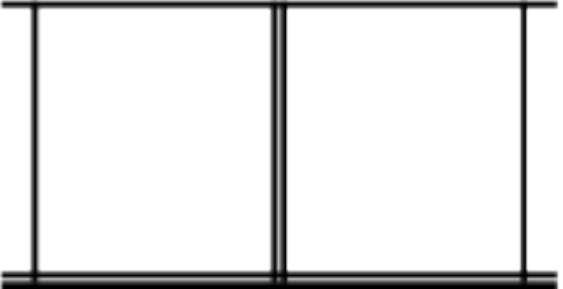}
    \caption{
      \label{fig::fire_ex}
      Integral family with three massive (double lines), three massless lines
      and an irreducible numerator (not shown).
      It it evaluated in forward-scattering kinematics, i.e., there is
      one external momentum, $p_1$, flowing through the upper massless line
      and another, $p_2$, through the lower massive lines.}
  \end{center}
\end{figure}

To demonstrate the use of {\tt FIRE} let us, for example, consider the
integral family of Fig.~\ref{fig::fire_ex} which has three massive internal
lines (with mass $m$).  For the external momenta we have $p_4=p_1$, $p_3=p_2$
with $p_1^2=p_2^2=0$.  Integrals of that type contribute to the
next-to-leading order corrections to double-Higgs boson production. In fact,
the imaginary part, which is a function of $x=m^2/s$ (with $s=(p_1+p_2)^2$),
is related to the total cross section via the optical theorem.  The input for
the {\tt Mathematica} part of {\tt FIRE} contains the following elements (for
a detailed description of the commands we refer to
Ref.~\cite{Smirnov:2014hma}):
\begin{verbatim}
(* load FIRE: *)
Get["FIRE5.m"];

(* define integral family: *)
Propagators = {m^2 - (v1-v2)^2, 
  m^2 - (p2-v1)^2, m^2 - (p2-v2)^2,
  -v2^2, -v1^2, -(p1+v1)^2, -(p1+v2)^2};
Internal = {v1,v2};
External = {p1,p2};

(* IBP relations: *)
PrepareIBP[];
kinset = {p1^2 -> 0, p2^2 -> 0, 
          p1*p2 -> s/2};
set1 = Internal;
set2 = Join[Internal,External];
ncount = 0;
startinglist = {};
For[ii=1,ii<=Length[set1],ii++,
    For[jj=1,jj<=Length[set2],jj++,
        ncount = ncount + 1;
        ff[ncount] = 
          IBP[set1[[ii]], set2[[jj]]
            ] /. kinset;
        startinglist = 
          Join[startinglist,{ff[ncount]}];
        ];
];

(* boundary conditions: only contributions 
   with cuts through at least 2 Higgs 
   lines are kept: *)
(RESTRICTIONS = { {0,-1,0,0,0,-1,0},
  {0,0,-1,0,0,0,-1},{0,0,0,0,0,-1,-1},
  {-1,0,0,0,0,0,0},{-1,-1,0,0,0,0,0},
  {-1,0,-1,0,0,0,0},{0,-1,-1,0,0,0,0} });
SYMMETRIES = { {1,3,2,5,4,7,6} };
Prepare[];

(* save data to top2l2h1a.start: *)
SaveStart["top2l2h1a"];
\end{verbatim}
The last command writes all generated information into the so-called ``start'' file
which, together with the list of integrals, serves as input for the reduction
step. The steering file, {\tt top2l2h1a.config}, 
for the latter has the following form
\begin{verbatim}
#threads    4
#variables  d,s,m
#start
#problem    1|7|top2l2h1a.start
#integrals  top2l2h1a.ind
#output     top2l2h1a.tab
\end{verbatim}
where we refer to Ref.~\cite{Smirnov:2014hma} for the precise meaning of the individual
commands. The integrals which shall be reduced can be found in
the file {\tt top2l2h1a.ind} which might have the form
\begin{verbatim}
{{1, {1, 1, 1, 2, 2, 2, 2}}, 
 {1, {1, 1, 1, 1, 1, 1, 2}},  
 {1, {1, 1, 1, 1, 1, 1, -1}}}
\end{verbatim}
Here the individual entries are lists where the integer in the first
entry numbers the family and the second entry contains seven integers
specifying the indices of the propagators as specified above (see
``\verb|Propagators|'').
The reduction is initiated with the help
of {\verb|./FIRE5 -c top2l2h1a|}. After the job is completed the reduction table
can be found in the file {\tt top2l2h1a.tab}
which can be read using again a {\tt Mathematica} session of {\tt FIRE}.

There are several benchmark calculations which have been performed with the
help of {\tt FIRE}. Among them is the reduction of all three-loop integrals
needed for the static
potential~\cite{Smirnov:2008pn,Smirnov:2009fh,Anzai:2009tm} which 
involves eight indices for massless relativistic propagators and in addition
three indices for static propagators of the form $1/k_0$. A particular
challenge poses the case for general QCD gauge parameter $\xi$ which involves 
about 20 million integrals, 60 times as much as the $\xi=0$ case. A further reduction
problem involves four-loop on-shell integrals needed for the relation between
the $\overline{\rm MS}$ and on-shell quark mass relation or the electron
anomalous magnetic moment (see, e.g., Ref.~\cite{Lee:2013sx}).

\subsection{Numerical evaluation of master integrals with {\tt FIESTA}}

{\tt FIESTA}~\cite{Smirnov:2008py,Smirnov:2009pb,Smirnov:2013eza} stands for
Feynman Integral Evaluation by a Sector decomposiTion Approach and is a
convenient tool to numerically evaluate Feynman integrals using the method of
sector decomposition. The latter is an algorithmic procedure to extract the
$\epsilon$ poles of a given Feynman integral in the so-called
alpha-representation and provide an integral representation for the
coefficients.  After the pioneering work of Binoth and
Heinrich~\cite{Binoth:2003ak,Binoth:2004jv} several programs have been
published where different strategies have been implemented. Among them are
{\tt sector\_decomposition}~\cite{Bogner:2007cr}, {\tt
  secdec}~\cite{Binoth:2004jv,Borowka:2012yc,Borowka:2013cma}, and {\tt
  FIESTA}~\cite{Smirnov:2008py,Smirnov:2009pb,Smirnov:2013eza}.

The basic philosophy of {\tt FIESTA} is that all kinematic variables are
specified at an early stage which is different from other approaches like,
e.g., {\tt secdec}, where generic manipulations are performed up to a certain
point and only then numerical values for masses and momenta are specified.

The use of {\tt FIESTA} splits into the following two steps:
In a first step the momentum integrals are transformed into the
alpha-representation and the sector decomposition algorithm is applied.  The
corresponding manipulations are performed in {\tt Mathematica} and can be done
in parallel mode.
For many applications this step is quite fast, however, quite often, in
particular at higher loop order, huge expressions are generated which require
main memory in the range of hundred Gigabyte. In such cases it is
convenient to store the results into a database~\cite{kyotocabinet} since in
general this step has to be performed only once.

The second step is concerned with the numerical integration.  In principle
this can also be performed within {\tt Mathematica}, which is advantageous for
small problems or during the developing phase of the program. Complicated problems
have to be integrated with the help of a {\tt C++} integrator which is based
on the {\tt Cuba} library~\cite{Hahn:2004fe,cuba}. It uses the expressions
stored in the database during step one which provides several advantages. For
example, it is possible to perform various runs choosing different values for
the number of points used for the integrations. Furthermore, it is possible to
copy the output of step one to a platform which is suitable for the numerical
integration in massive parallel mode.

\begin{figure}[t]
  \begin{center}
    \includegraphics[width=.8\linewidth]{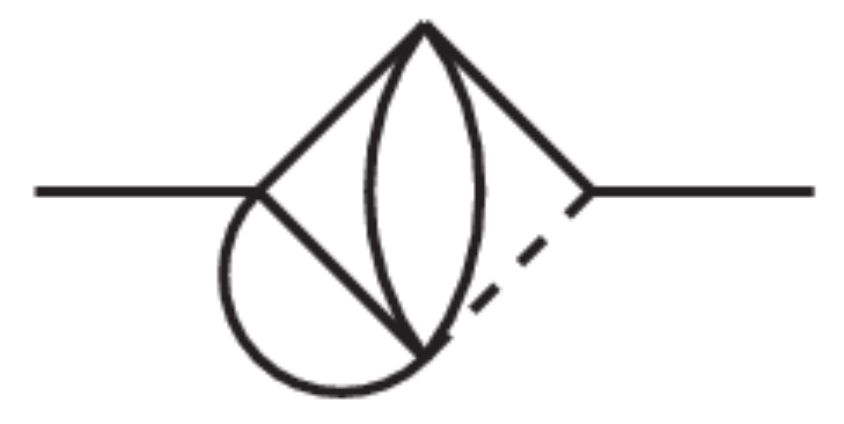}
    \caption{\label{fig::diag_4los}
      Sample on-shell Feynman diagram where solid and dashed lines denote massive
      and massless lines, respectively.
    }
  \end{center}
\end{figure}

Let us as an example consider the Feynman diagram in Fig.~\ref{fig::diag_4los}
which enters the four-loop relation between the $\overline{\rm MS}$-on-shell
quark mass. Executing the {\tt Mathematica} file
\begin{verbatim}
Get["FIESTA3.m"];
NumberOfSubkernels=8;
NumberOfLinks=8;
UsingC=True;
UsingQLink=True;
ComplexMode=False;
SDEvaluate[ UF[ {k1,k2,k3,k4},
                {-(k1+q1)^2+m^2,
                 -(k3+q1)^2+m^2,
                 -(k1-k2)^2+m^2,
                 -(k2-k3)^2+m^2,
                 -(k1-k4)^2+m^2,
                 -k4^2+m^2,
                 -k3^2},
                {m->1,q1^2->1}],
                {1,1,1,1,1,1,1},6]
\end{verbatim}
prepares both the integrand and performs the numerical integration using
the corresponding {\tt C} routines in the background. The result
which is printed on the screen reads
\begin{verbatim}
-276.907674 - 0.625006/ep^4 - 
 4.937615/ep^3 + (-24.441689 + 
 0.002*pm69)/ep^2 + (-85.919995 + 
 0.015937*pm70)/ep + 0.083469*pm71 + 
 ep*(-864.271585 + 0.468742*pm72) + 
 ep^2*(-1503.357843 + 2.093833*pm73) + 
 ep^3*(-6224.681821 + 9.755544*pm74) + 
 ep^4*(11328.088699 + 40.591518*pm75) + 
 ep^5*(-18622.607506 + 176.767061*pm76) + 
 ep^6*(537473.776134 + 713.790523*pm77)
\end{verbatim}
The symbols \verb|pm| indicate the uncertainty 
due to the Monte Carlo integration.
In case the option \verb|OnlyPrepare = True;| is added 
to the {\tt Mathematica} file the integrand is prepared 
and stored to disk. Furthermore the command 
is printed on screen which invokes the 
numerical integration from the shell without
reference to {\tt Mathematica}.

\begin{figure}[t]
  \begin{center}
    \includegraphics[width=1.\linewidth]{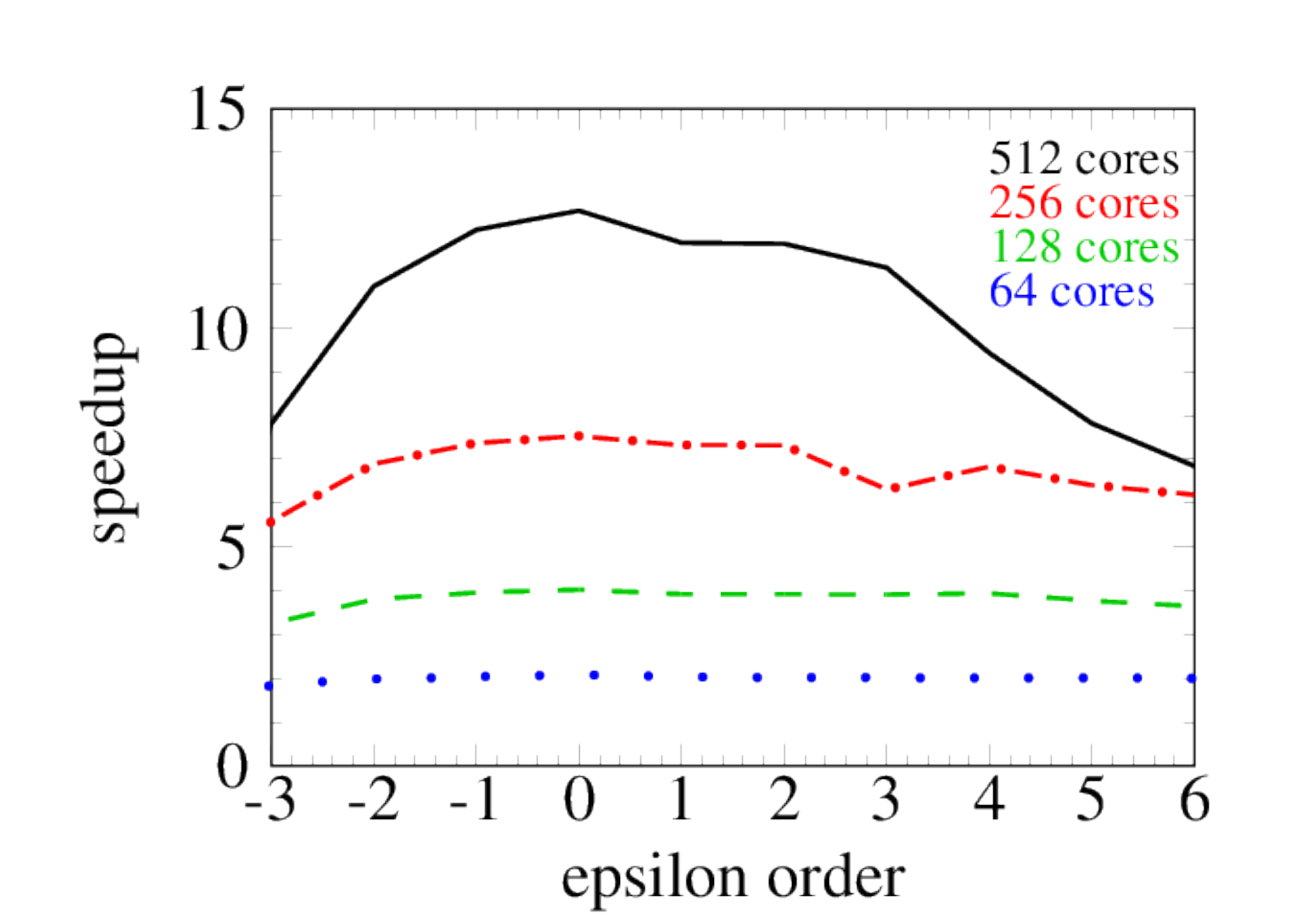}
    \caption{\label{fig::FIESTA_speedup}
      Speedup of the calculation of the various $\epsilon$ orders
      of Feynman diagram given in Fig.~\ref{fig::diag_4los} using
      64 (dotted), 128 (dashed), 256 (dash-dotted) and 512 (solid) cores 
      normalized to the 32-core run.
    }
  \end{center}
\end{figure}

The result from runs performed at the High Performance Computing Center
Stuttgart (HLRS) are shown in Fig.~\ref{fig::FIESTA_speedup} where the speedup
for the individual $\epsilon^n$ terms ($n=-3,\ldots,6$) is shown.  The blue
(dotted), green (dashed), red (dash-dotted) and black (solid) curves (from
bottom to top) corresponds to the use of 64, 128, 256 and 512 cores where the
results have been normalized to the 32-core run. It is interesting to note
that an ideal speedup is obtained for 64 cores. Also for 128 cores the curve
is close to the maximal value of 4. Using 256 instead of 32 cores still shows a
quite flat behaviour with a speedup between 6 and 7.  Strong variations in the
speedup are observed for the use of 512 cores.  The relatively low value for
$1/\epsilon^{-3}$ can be explained with the fact that probably the expression,
which shall be integrated, is too simple.  On the other hand, for the
(complicated) expression of the $\epsilon^{6}$ coefficient it might be that
the disk access becomes the bottle neck.

The main purpose of {\tt FIESTA} is the fast and convenient cross check of
analytic calculations. Within CRC/TR~9 is has been applied in this way 
to several problems. An early version of {\tt FIESTA} has been used to cross
check the master integrals which contribute to the three-loop static
potential~\cite{Smirnov:2008pn,Smirnov:2009fh,Anzai:2009tm}. Furthermore
thirteen four-loop on-shell integrals 
contributing to the $\overline{\rm MS}$-on-shell quark mass relation and to
the muon anomalous magnetic moment, which have been computed analytically in
Ref.~\cite{Lee:2013sx}, have been cross-checked numerically with {\tt FIESTA}.
Recently also analytic results for master integrals 
of double-box topologies in the physical region have been cross-checked with
the help of {\tt FIESTA}~\cite{Caola:2014lpa}.

There are also several projects where {\tt FIESTA} has been used to evaluate
the most complicated or even the major part of the master integrals
numerically. For example, in the first calculation of the three-loop
corrections of the quark and gluon form factor~\cite{Baikov:2009bg} (see also
Ref.~\cite{Gehrmann:2010ue}) one coefficient in the $\epsilon$ expansion of
the three most complicated integrals could not be evaluated
analytically. Thus, the numerical results of {\tt FIESTA} have been used
which, for all practical purposes, leads to final results with sufficient
precision. The analytic calculation of the missing master integrals has been
performed in Ref.~\cite{Lee:2010ik} and perfect agreement with the numerical
result has been found.

For the calculation of the three-loop matching coefficient between QCD and
non-relativistic QCD (NRQCD) of the vector current~\cite{Marquard:2014pea}
even the majority of the about 100 master integrals have been computed
numerically with the help of {\tt FIESTA}. In such cases it is important to
perform strong cross checks. Among them are the change of the parametrization
of the individual integrals. Thus, in intermediate steps different expressions
are generated which are then integrated numerically. Furthermore, it is
possible to choose a different integrals basis and evaluate the new integrals
again with the help of {\tt FIESTA}. The agreement of the final expression
within the numerical uncertainty among the two set of master integrals serves
as a strong checked for the applicability of {\tt FIESTA}.

%- }}}

%- {{{ Summary

\section{\label{sec::sum}Summary}

The computer algebra program \form{} is designed to handle huge expressions 
in a quite effective way. Still, for some physical applications even 
\form{} would take several years which make a practical calculation 
impossible.

In the recent years parallel versions of \form{}, \parform{} and \tform{},
have been developed and in the meantime they have become a reliable tools to
perform computer algebra in parallel. \parform{} has demonstrated a good
speedup behaviour both on SMP computers and on different cluster
architectures.  Furthermore, for the current version of \parform{} the \form{}
programs written for the sequential version need not to be modified.

\tform{} is a parallel version of \form{} based on {\tt POSIX} threads 
and thus is bound to run on a single node. However, there is less overhead
connected to the parallelization and thus \tform{} shows a slightly better
performance than \parform{}. 

The main advantage of using a parallel version of \form{} is the reduction of
the wall clock time. In fact, there are a number of calculations where it has
been exploited that a speedup of about 10 can be reached with 16 cores
and thus the result was available after about a month instead of a year.
A further advantage of using \tform{} or \parform{} is the fact that
the size of the intermediate results, which have to be handled by the
individual CPU, is smaller since the workload is
distributed among several workers. This advantage becomes particularly
evident when using \parform{} on a cluster. In that case the intermediate 
expressions are stored into files which are located on different nodes.

To obtain an even better speedup behaviour it would be necessary
to improve the slope of the speedup curves and to push the
flattening to higher number of processors. One starting point which could help to
improve the situation is the sorting procedure.
Another idea might be the combination of \parform{} and \tform{} 
which could be an ideal tool for a cluster with multi-core nodes.

In this article we also describe the programs {\tt FIRE} and {\tt FIESTA}.
{\tt FIRE} can be used for the reduction of integrals belonging to a given 
integral family to master integrals. {\tt FIESTA}, on the other hand
is a user-friendly tool to numerically compute the coefficients of the
$\epsilon$ expansion of multi-loop integrals.

%- }}}

\section*{Acknowledgements}
This work is supported by the Deutsche
Forschungsgemeinschaft in the Sonderforschungsbereich Transregio~9
``Computational Particle Physics''.
We acknowledge the use of the High Performance Computing Center Stuttgart
(HLRS) where part of the calculations connected to {\tt FIESTA} have
been carried out. In this context we also acknowledge the help of Peter
Marquard. 

%- {{{ Appendix:

%% The Appendices part is started with the command \appendix;
%% appendix sections are then done as normal sections
%\appendix

%\section{????}
%% \label{}

%- }}}

%- {{{ References:
%%
%% Following citation commands can be used in the body text:
%% Usage of \cite is as follows:
%%   \cite{key}         ==>>  [#]
%%   \cite[chap. 2]{key} ==>> [#, chap. 2]
%%

%% References with BibTeX database:
%\nocite{*}
%\bibliographystyle{elsarticle-num}
%\bibliography{martin}

%% Authors are advised to use a BibTeX database file for their reference list.
%% The provided style file elsarticle-num.bst formats references in the required Procedia style

%\section*{References}

%% For references without a BibTeX database:

%- }}}

\end{document}